\documentclass[a4paper, times, 10pt,twocolumn]{article}
\usepackage[top=4.9cm,bottom=3.7cm,left=1.5cm,right=1.5cm]{geometry}
\usepackage{ICMLC}
\usepackage{times}
\usepackage{graphicx}
\usepackage{indentfirst}
\usepackage{latexsym}

\usepackage{threeparttable}
\usepackage{multirow}
\usepackage[margin=8pt,font=footnotesize,labelfont=bf,labelsep=period
]{caption}

\pdfpagewidth=\paperwidth
\pdfpageheight=\paperheight
\begin{document}

\title{DATA-DRIVEN APPROACH FOR QUALITY EVALUATION ON KNOWLEDGE SHARING PLATFORM}  

\author{\bf{\normalsize{LU XU${^1}$, JINHAI XIANG${^1}$, YATING WANG${^2}$, FUCHUAN NI${^1}{^*}$}}\\ 
\\
\normalsize{$^1$College of Informatics, Huazhong Agricultural University, No.1, Shizishan Street, Hongshan District, Wuhan, China}\\
\normalsize{$^2$School of Economics and Management, Hubei University of Technology, No.28, Nanli Road, Hongshan District, Wuhan, China} \\
\normalsize{E-MAIL: xulu\_coi@webmail.hzau.edu.cn, fcni\_cn@mail.hzau.edu.cn$^{*}$}\\
\\}

\maketitle 
\thispagestyle{empty}

\begin{abstract}
   {In recent years, voice knowledge sharing and question answering (Q\&A) platforms have
   	attracted much attention, which greatly facilitate the knowledge
   	acquisition for people.
   	However, little research has evaluated on the quality evaluation on voice knowledge sharing.
   	This paper presents a data-driven approach to automatically evaluate the quality of a specific Q\&A platform (Zhihu Live).
   	Extensive experiments demonstrate the effectiveness of the proposed method. Furthermore, we introduce
   	a dataset of Zhihu Live as an open resource for researchers in related areas.
   	This dataset will facilitate the development of new methods on knowledge sharing services quality evaluation.}
\end{abstract}


\section{Introduction}
Knowledge sharing platforms such as Quora\footnote{https://www.quora.com/} and 
Zhihu\footnote{https://www.zhihu.com/} emerge as very convenient tools for acquiring knowledge. These question and answer (Q\&A) platforms are newly emerged communities about knowledge acquisition, experience sharing and social networks services (SNS).

Unlike many other Q\&A platforms, Zhihu platform resembles a social network community.
Users can follow other people, post  ideas, up-vote or down-vote answers, and write their own answers.
Zhihu allows users to keep track of specific fields by following related topics, 
such as ``Education'', ``Movie'', ``Technology'' and ``Music''.
Once a Zhihu user starts to follow a specific topic or a person, the related updates are automatically pushed to the user's feed timeline.

\begin{figure}[!thbp]
	\centering
	\includegraphics[scale=0.28]{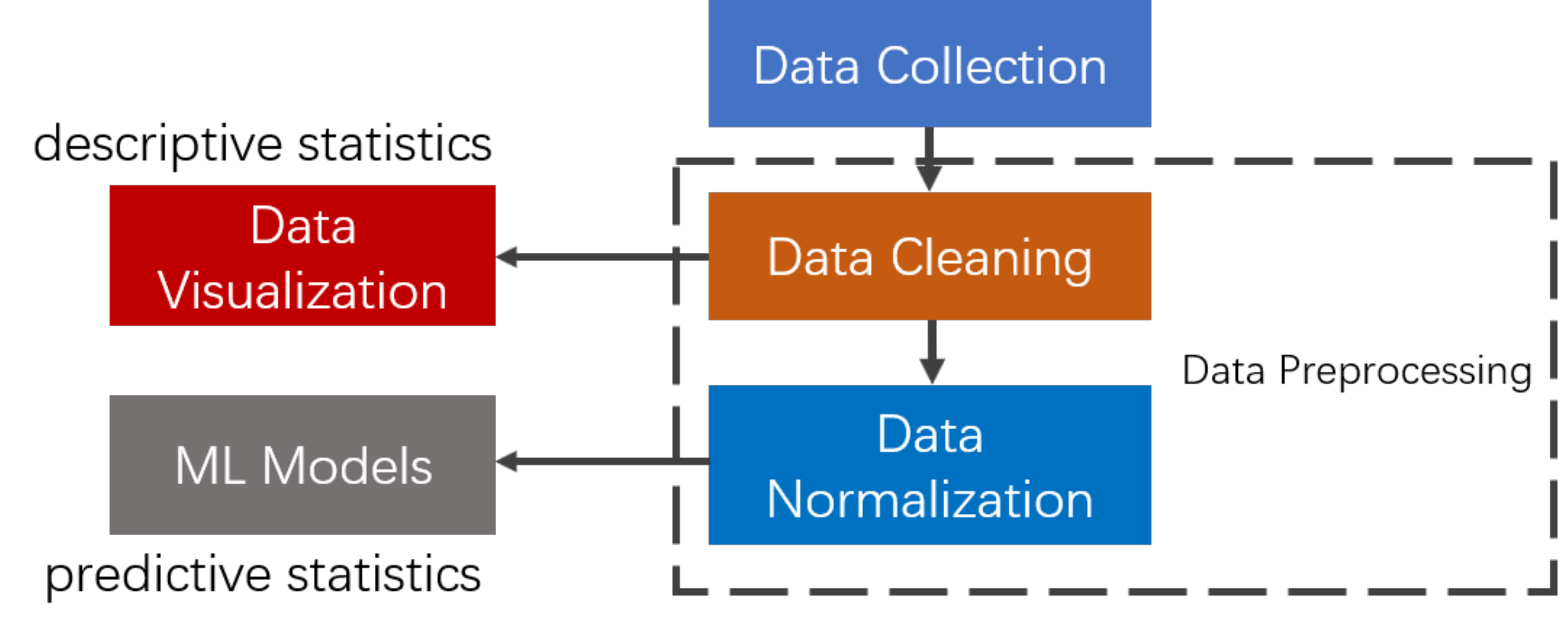}
	\caption{
		The pipeline of our data-driven method.
	}
	\label{fig:our_method_pipeline}
\end{figure}

Although these platforms have exploded in popularity, they face some potential problems.
The key problem is that as the number of users grows, a large volume of low-quality questions and answers emerge 
and overwhelm users, which make users hard to find relevant and helpful information.

Zhihu Live is a real-time voice-answering product on the Zhihu platform, which enables the
speakers to share knowledge, experience, and opinions on a subject.
The audience can ask questions and get answers from the speakers as well.
It allows communication with the speakers easily and efficiently through the Internet.
Zhihu Live provides an extremely useful reward mechanism (like up-votes, following growth and economic returns), to encourage high-quality content providers to generate high-level information on Zhihu platform.

However, due to the lack of efficient filter mechanism and evaluation schemes, 
many users suffer from lots of low-quality contents, which affects the service negatively.
Recently, studies on social Q\&A platforms and knowledge sharing are rising and have achieved many promising results.
Shah et al. \cite{Shah2010Evaluating} propose a data-driven approach with logistic regression and carefully designed hand-crafted features to predict the answer
quality on Yahoo! Answers.
Wang et al. \cite{wang2013wisdom} illustrate that heterogeneity in the user and question graphs are
important contributors to the quality of Quora's knowledge base.
Paul et al. \cite{Paul2012Who} explore reputation mechanism in quora through detailed data analysis, their experiments indicate that social voting helps users identify and promote good content but is prone to preferential attachment.
Patil et al. \cite{patil2016detecting} propose a method to detect experts on Quora by their activity,
quality of answers, linguistic characteristics and temporal behaviors, and achieves 97\% accuracy and 0.987 AUC.
Rughinis et al. \cite{Rughinis2014Computer} indicate that there are different regimes of engagement at the 
intersection of the technological infrastructure and users' participation in Quora.

All of these works are mainly focused on answer ranking and answer quality evaluation.
But there is little research achievement about quality evaluation in voice-answering areas.
In this work, we present a data-driven approach for quality evaluation about Zhihu Live,
by consuming the dataset we collected to gather knowledge and insightful conclusion.
The proposed data-driven approach includes data collection, storage, preprocessing, data analysis, and predictive analysis via machine learning. The architecture of our data-driven method is shown in Fig.~\ref{fig:our_method_pipeline}.
The records are crawled from Zhihu Live official website and stored in MongoDB. Data preprocessing methods include cleaning and data normalization to make the dataset satisfy our target problem. Descriptive data analysis and predictive analysis are also conducted for deeper analysis about this dataset.

The main contributions of this paper are as follows: (1) We release a public benchmark dataset which contains 7242 records and 286,938 text comments about Zhihu Live. Detailed analysis about the dataset is also discussed in this paper. This dataset could help researchers verify their ideas in related fields. (2) By analyzing this dataset, we gain several insightful conclusion about Zhihu Live. (3) We also propose a multi-branched neural network (MTNet) to evaluate Zhihu Lives' scores. The superiority of our proposed model is demonstrated by comparing performance with other mainstream regressors.

The rest of this paper is organized as follows:
Section 2 describes detailed procedures of ZhihuLive-DB collection,
and descriptive analysis. Section 3 illustrates our proposed MTNet. 
In section 4, we give a detailed description of experiments,
and the last section discusses the conclusion of this paper and future work.

\section{Zhihu Live Dataset}
\subsection{Data Collection}
In order to make a detailed analysis about Zhihu Live with data-driven approach, 
the first step is to collect Zhihu Live data.
Since there is no public dataset available for research and no official APIs,
we develop a web spider with python requests library\footnote{http://docs.python-requests.org/en/master/} 
to crawl data from Zhihu Live official website\footnote{https://www.zhihu.com/lives}.
Our crawling strategy is breadth-first traverse (we crawl the records one by one from the given URLs,
and then extract more detailed information from sub URLs).
We follow the crawler-etiquette defined in Zhihu's robots.txt. So we randomly set 2 to 5 seconds pause after per crawling 
to prevent from being banned by Zhihu, and avoid generating abnormal traffic as well.
Our spider crawls 7242 records in total. Majority of the data are embedded in Ajax calls. In addition, we also crawl 286,938 comments of these Zhihu Lives. All of the datasets are stored in \emph{MongoDB}, a widely-used NoSQL database.

\subsection{Statistical Analysis}
The rating scores are within a range of $[0, 5]$. We calculate \emph{min}, \emph{Q1}, \emph{median}, \emph{Q3}, \emph{max}, \emph{mean}, and \emph{mode} about \textit{review count} (see Table~\ref{tb:review_count}). Because the number of received review may greatly influence the reliability of the review score.
From Table~\ref{tb:review_count} we can see that many responses on Zhihu Live receive no review at all, which are useless for quality evaluation.

\begin{table}[!ht]
	\renewcommand{\arraystretch}{1.3}
	\caption{Review count of ZhihuLive-DB.}
	\label{tb:review_count}
	\centering
	\begin{tabular}{|c|c|c|c|c|c|c|}
		\hline
		\textbf{Min} & \textbf{Q1} & \textbf{Median} & \textbf{Q3}
		& \textbf{Max} & \textbf{Mean} & \textbf{Mode}\\
		\hline
		0 & 11 & 35 & 99 & 22539 & 119 & 3 \\
		\hline
	\end{tabular}
\end{table}

One of the most challenging problems is no unique standard to evaluate a Zhihu Live
as a low-quality or high-quality one. A collection of people may highly praise a Zhihu Live while others may not.
In order to remove the sample bias, we delete those records whose
review count is less than Q1 (11). So we get 5477 records which belong to 18 different fields.

The statistics of review scores after deletion are shown in Table~\ref{tb:revire_score statistic}. The mean score of 5477 records is 4.51, and the variance is 0.16. It indicates that the majority of Zhihu Lives are of high quality, and the users' scores are relatively stable.

\begin{table}[!ht]
	\renewcommand{\arraystretch}{1.3}
	\caption{Statistics of review scores after deletion.}
	\label{tb:revire_score statistic}
	\centering
	\begin{tabular}{|c|c|c|c|}
		\hline
		\textbf{Mean} & \textbf{Mode} & \textbf{Median} & \textbf{Variance} \\
		\hline
		4.51 & 4.67 & 4.6 & 0.16 \\
		\hline
	\end{tabular}
\end{table}

\begin{table}[!ht]
	\caption{Influence of badges on Zhihu Live scores.}
	\label{tb:badge_scores}
	\begin{tabular}{|c|c|c|c|}
		\hline
		Badges\_Num & Speakers\_Num & Avg\_Score & Median\_Score\\
		\hline
		2 & 446 & 4.55 & 4.63\\
		1 & 1475 & 4.51 & 4.60\\
		0 & 3286 & 4.50 & 4.60\\
		\hline
	\end{tabular}
\end{table}

\emph{Badge} in Zhihu represents identity authentication of public figures
and high-quality answerers. Only those who hold a Ph.D. degree or experts in a specific
domain can be granted a badge. Hence, these speakers tend to host high-quality Zhihu Lives theoretically.
Table~\ref{tb:badge_scores} shows that 3286 speakers hold no badge, 1475 speakers hold 1 badge, and 446 speakers hold 2 badges, respectively. The average score of Zhihu Lives given by two badges holders is slightly higher than others.
We can conclude that whether the speaker holds badges does have slightly influence on the Zhihu Live quality ratings, which is consistent with our supposition.

Furthermore, we calculate the average scores of different Zhihu Live types (See Table~\ref{tb:type_avg_score}).
We find that \emph{Others}, \emph{Art} and \emph{Sports} fields contain more high-quality Zhihu Lives,
while \emph{Delicacy}, \emph{Business} and \emph{Psychology} fields contain more low-quality Lives.
We can conclude that the topics related to self-improvement tend to receive more positive comments.

\begin{table}[!htpb]
	\renewcommand{\arraystretch}{1.3}
	\caption{Average scores of different Zhihu live types.}
	\label{tb:type_avg_score}
	\centering
	\begin{tabular}{|c|c|c|c|}
		\hline
		\textbf{Type} & \textbf{Avg Score} & \textbf{Type} & \textbf{Avg Score} \\
		\hline
		Others & 4.78 & Design & 4.48 \\
		Art & 4.64 & Finance & 4.48 \\
		Health & 4.61 & Career & 4.46 \\
		Law & 4.59 & Internet & 4.45 \\
		Education & 4.59 & Lifestyle & 4.44 \\
		Sports & 4.57 & Tour & 4.44 \\
		Technology & 4.54 & Business & 4.43 \\
		Reading & 4.53 & Psychology & 4.37 \\
		Amusement & 4.52 & Delicacy & 4.34 \\
		\hline
	\end{tabular}
\end{table}

There are two types of Zhihu accounts: personal and organization. From Table~\ref{tb:gender_and_type}, we can see that the majority of the Zhihu Live speakers are men with personal accounts. Organizations are less likely to give presentation and share ideas upon Zhihu Live platform.

\begin{table}[htbp]
	\centering
	\caption{Statistical information of speakers's gender and type.}
	\label{tb:gender_and_type}
	\begin{tabular}{|c|c|c|c|}
		\hline
		\textbf{Gender} & \textbf{Number} & \textbf{Type} & \textbf{Number}\\
		\hline
		Male & 4263 & Personal & 5349\\
		Female & 1214 & Organization & 128\\
		\hline
	\end{tabular}
\end{table}

\subsection{Comments Text Analysis}
\begin{figure}[htbp]
	\centering
	\includegraphics[width=0.5\textwidth]{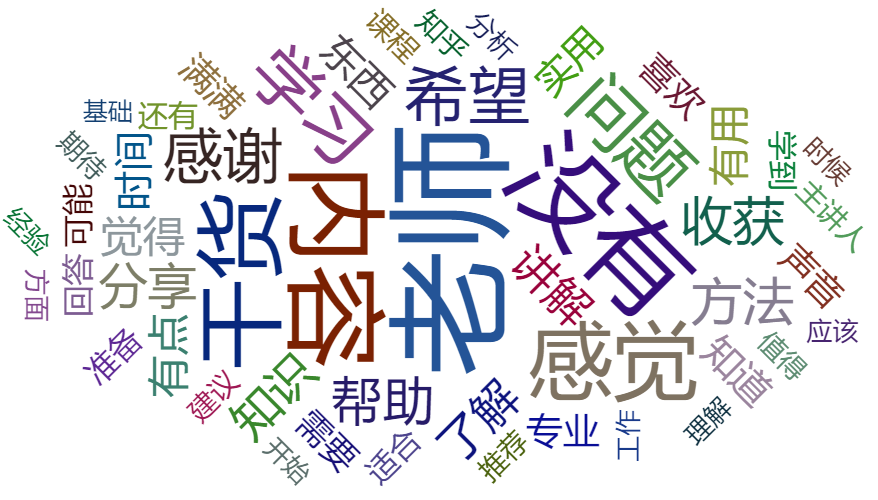}
	\caption{Hot words extraction via \textit{TextRank}. Bigger font denotes higher weight.}
	\label{fig:hot_words}
	
\end{figure}

Apart from analyzing Zhihu Live dataset, we also adopt \emph{TextRank}~\cite{mihalcea2004textrank} algorithm to calculate TOP-50 hot words with wordcloud visualization (see Fig.~\ref{fig:hot_words}). Bigger font denotes higher weight of the word, we can see clearly that the majority of the comments show contentment about Zhihu Lives, and the audience care more about ``content'', ``knowledge'' and ``speaker''.

\subsection{Performance Metric}
We define the quality evaluation problem as a standard regression task since the scores we aim to predict are continuous values.
Hence we use \emph{Mean Absolute Error (MAE)} and \emph{Root Mean Square Error (RMSE)}
to estimate the performance of diverse learning algorithms.
MAE and RMSE are used to evaluate the fit quality of the learning algorithms,
if they are close to zero, it means the learning algorithm fits the dataset well.

\begin{equation}
RMSE=\sqrt{\frac{1}{m}\sum_{i=1}^m\big(h(x^{(i)})-y^{(i)})^2}, 
\end{equation}

\begin{equation}
MAE=\frac{1}{m} \sum_{i=1}^m|h(x^{(i)})-y^{(i)}|, 
\end{equation}
where $\emph{m}$ denotes the number of samples, $x^{(i)}$ denotes the
input feature vector of a sample $i$, $h(\bullet)$ denotes the learning algorithm,
$y^{(i)}$ denotes the groundtruth score of a Zhihu Live response $i$.

The results are calculated by randomly selecting 80\% in the dataset as training set,
and the remaining records as test set.

\section{MTNet}
In this section, we first give a brief introduction of the neural network and then present a description of our proposed 
MTNet to predict the quality of responses in detail.

\subsection{Deep Neural Network}
Deep neural network (DNN) has aroused dramatically attention due to their extraordinary performance
in computer vision \cite{Krizhevsky2012ImageNet,szegedy2015going}, speech recognition~\cite{Goodfellow-et-al-2016} and natural language
processing (NLP) \cite{zhang2018deep} tasks. We apply DNN to our Zhihu Live quality evaluation problem aiming to approximate a function $f^*$ which can accurately predict a Zhihu Live's score.

In our quality evaluation task, we take multiple layer perception~\cite{Goodfellow-et-al-2016} as the basic composition block of MTNet. Since we treat the Zhihu Live quality evaluation problem as a regression task, we set the output neuron equal to 1. DNNs are trained by backpropagation algorithm~\cite{Goodfellow-et-al-2016}.

The calculation details of neural network can be illustrated as: 
\begin{equation}
f(W^tx)=\sigma(\sum_{i=1}^nW_ix_i+b)
\end{equation}
where $f(W^tx)$ represents output of a neuron, $W_i$ represents weights of the connections, $b$ represents bias, $\sigma$ represents 
nonlinear activation function (sigmoid, tanh and ReLU are often used in practice).

\subsection{MTNet Architecture}
\begin{figure}[!htpb]
	\centering
	\includegraphics[width=0.4\textwidth]{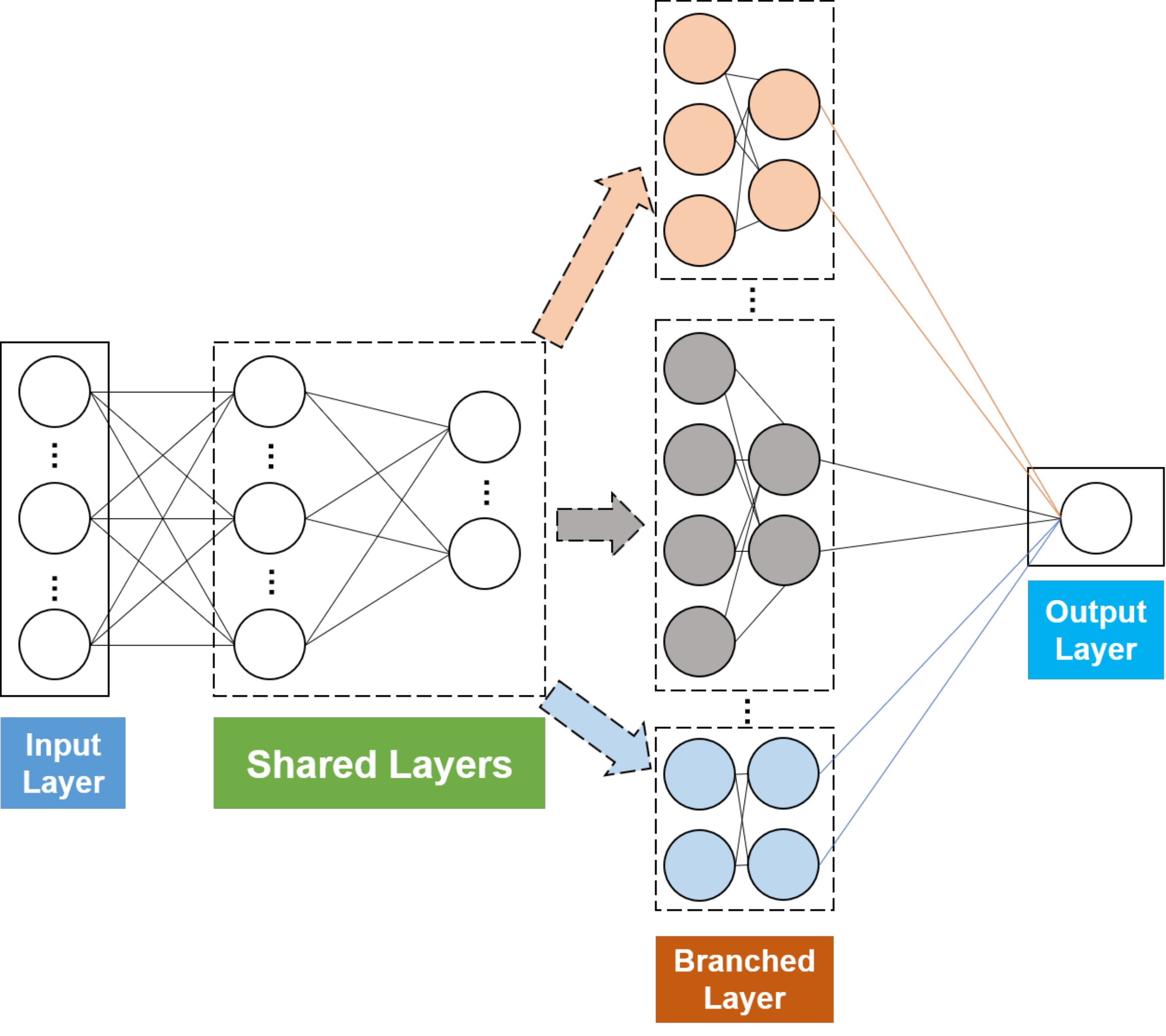}
	\caption{Overall architecture of MTNet.}
	\label{fig:mtb_dnns}
\end{figure}

The architecture of our proposed MTNet is shown in Fig.~\ref{fig:mtb_dnns}. It includes 4 parts: an input layer for receiving raw data; shared layers for general feature extraction through stacked layers and non-linear transformation; branched layers for specific feature extraction; and the output layer with one neuron. The output of the last shared layer is fed into different branches. These branches are trained jointly.
In the last shared layer, the information flow is split into many branches~\cite{szegedy2015going}, which enables feature sharing and reuse. Finally, the output result is calculated in the output layer by averaging outputs from these branches\cite{ciregan2012multi}. The overall neural network with different branches is trained in parallel. The detailed configuration of MTNet is listed in Tabel~\ref{tab:mtnet}.

\begin{table}[htbp]
	\centering
	\caption{Detailed configuration of MTNet.}
	\label{tab:mtnet}
	\begin{tabular}{|c|c|}
		\hline
		\textbf{Layer}			         & \textbf{Neurons and Layer Depth}\\ \hline
		Input Layer                      & 23     \\ \hline
		Shared Layers                    & 16-8-8 \\ \hline
		\multirow{3}{*}{Branched Layers} & 8-5-1  \\ \cline{2-2} 
		& 8-4-1  \\ \cline{2-2} 
		& 8-3-1  \\ \hline
		Output Layer                     & 1      \\ \hline
	\end{tabular}
\end{table}

The advantages of MTNet are as follows:
\begin{itemize}
	\item With multi-branched layers, different data under diverse levels can be fed into different branches, which enables MTNet extract multi-level features for later regression.
	\item Multi-branched architecture in our MTNet can also act as an ensemble method~\cite{ciregan2012multi}, which promotes the performance as well.
\end{itemize}

We use mean square error (MSE) with $L_2$ regularization as the cost function.
\begin{equation}
\mathcal{J}(\theta)=\frac{1}{N}\sum_{i=1}^N(f(x_i,\theta)-y_i)^2+\lambda||\theta||_2^2
\end{equation}
where $x_i$ denotes the raw input of $i$-th data sample, $N$ denotes the capacity of dataset, $y_i$ denotes groundtruth score of $i$-th Zhihu Live. $||\theta||_2^2$ denotes $L_2$ regularization to prevent from overfitting.

\section{Experiments}
We implement our method based on Scikit-Learn~\cite{scikit-learn} and PyTorch~\footnote{https://pytorch.org/}, and the experiments are conducted on a server with NVIDIA Tesla K80 GPU.

\subsection{Data Preprocessing}
Several features' types in ZhihuLive-DB are not numerical, while machine learning
predictor can only support numerical values as input. We clean the original dataset through the following preprocessing methods.
\begin{itemize}
	\item For categorical features, we replace them with one-hot-encoding~\cite{scikit-learn}. 
	\item The missing data is filled with \emph{Median} of each attribute.
	\item We normalize the numerical values with minimum subtraction and range division to ensure values [0, 1] intervals.
	\item The review scores are used as labels in our experiments, our task is to precisely estimate the scores with MTNet.
	Since the data-driven methods are based on crowd wisdom on Zhihu Live platform, they don't need any additional labeling work, and ensure the reliability of the scores of judgment as well.
\end{itemize}

\subsection{Feature Selection}
Since \emph{feature selection} plays an import part in a data mining task, conventional feature extraction methods need domain knowledge~\cite{guyon2003introduction}.
Feature selection influences model's performance dramatically \cite{Domingos2012A}.

For conventional regression algorithms, we conduct feature selection by adopting
the best \emph{Top K} features through univariate statistical tests. The hyper-parameter such as
regularization item $\lambda$ is determined through \emph{cross validation}.
For each regression algorithm mentioned above, the hyper-parameters are carefully tuned, and the
hyper-parameters with the best performance are denoted as the final comparison results.
The details of \emph{f\_regression}~\cite{Bishop2006Pattern,scikit-learn} feature selection are as follows:
\begin{itemize}
	\item We calculate the correlation between each regressor and label as:
	$((X[:, i] - mean(X[:, i])) * (y - mean_y)) / (std(X[:, i]) * std(y))$.
	\item We convert the correlation into an F score and then to a p-value.
	\item Finally, we get 15-dimension feature vector as the input for conventional (non-deep learning based) regressors.
\end{itemize}

Deep neural network can learn more abstract features via stacked layers. Deep learning has empowered many AI tasks
(like computer vision \cite{Krizhevsky2012ImageNet} and natural language processing \cite{zhang2018deep}) in an end-to-end fashion. We apply deep learning to our Zhihu Live quality evaluation problem.
Furthermore, we also compare our MTNet algorithm with baseline models with carefully designed features.

\subsection{Experimental Results}
We train our MTNet with Adam optimizer for 20 epochs. We set batch size as 8, and weight decay as 1e-5, we adopt 3 branched layers in MTNet. Detailed configuration is shown in Table~\ref{tab:mtnet}.
We use \textit{ReLU} in shared layers, and \textit{relu6} in branched layers to prevent information loss. Our proposed MTNet achieves 0.2250 on MAE and 0.3216 on RMSE, respectively.

We compare MTNet with other mainstream regression algorithms~\cite{Bishop2006Pattern} (linear regression, KNN, SVR, Random Forest and MLP). The architecture of MLP is 15-16-8-8-1, where each number represents the number of neurons in each layer. We try three kinds of kernels (RBF kernel, linear kernel, and poly kernel) with SVR in our experiments for fair comparison.

\begin{table}[hbtp]
	\renewcommand{\arraystretch}{1.3}
	\caption{Performance comparison with other regression models. The best results are given in bold style.}
	\label{tb:regression performance comparison}
	\centering
	\begin{tabular}{|c|c|c|}
		\hline
		\textbf{Regressor} & \textbf{MAE} & \textbf{RMSE} \\
		\hline
		Linear Regression & 0.2366 & 0.3229 \\ \hline
		KNN Regression & 0.2401	& 0.3275 \\ \hline
		SVR (RBF) & 0.2252 & 0.3270 \\ \hline
		SVR (Linear) & 0.2257 & 0.3267 \\ \hline
		SVR (Poly) & 0.2255	& 0.3268 \\ \hline
		Random Forest Regressor & 0.2267 & 0.3244\\\hline
		MLP & 0.2397 & 0.3276 \\ \hline
		MTNet & \textbf{0.2250} & \textbf{0.3216} \\ \hline
	\end{tabular}
\end{table}

The results are listed in Table~\ref{tb:regression performance comparison}.
Our method achieves the best performance in contrast to the compared baseline regressors.

\section{Conclusion}

In this paper, we adopt a data-driven approach which includes data collection, data cleaning, data normalization, descriptive analysis and predictive analysis, to evaluate the quality on Zhihu Live platform. To the best of our knowledge, we are the first to research quality evaluation of voice-answering products.
We publicize a dataset named ZhihuLive-DB, which contains 7242 records and 286,938 comments text for researchers to evaluate Zhihu Lives' quality.
We also make a detailed analysis to reveal inner insights about Zhihu Live. In addition, we propose MTNet to accurately predict Zhihu Lives' quality. Our proposed method achieves best performance compared with the baselines.

As knowledge sharing and Q\&A platforms continue to gain a greater popularity,
the released dataset ZhihuLive-DB could greatly help researchers in related fields.
However, current data and attributes are relatively unitary in ZhihuLive-DB. The malicious comment and assessment
on SNS platforms are also very important issues to be taken into consideration. In our future work, we will gather
richer dataset, and integrate malicious comments detector into our data-driven approach.

\section*{Acknowledgements}

Supported by Foundation Research Funds for the Central Universities
(Program No.2662017JC049) and State Scholarship Fund (NO.261606765054).


\end{document}